\documentclass[12pt,a4paper]{iopart}

\usepackage{iopams}  
\usepackage{graphicx}
\usepackage[breaklinks=true,colorlinks=true,linkcolor=blue,urlcolor=blue,citecolor=blue]{hyperref}
\usepackage{setspace}

\begin{document}

\title[Flow optimization process in a transportation network]{Flow optimization process in a transportation network}

\author{F. L. Forgerini}
\address{Institute of Humanities, Arts and Science\\
 Campus Jorge Amado, Universidade Federal do Sul da Bahia\\
45613-204 \hspace{5mm} Itabuna - BA \hspace{5mm} Brazil}
\ead{fabricio.forgerini@ufsb.edu.br}

\author{O. F. de Sousa}
\address{Centro de Forma\c{c}\~ao de Professores, Universidade Federal do Rec\^oncavo da Bahia\\
45300-000 \hspace{5mm} Amargosa - BA \hspace{5mm} Brazil}
\ead{orahcio@ufrb.edu.br}

\begin{abstract}
Numerous networks, such as transportation, distribution and delivery networks optimize their designs in order to increase efficiency and lower costs, improving the stability of its intended functions, etc. Networks that distribute goods, such as electricity, water, gas, telephone and data (Internet), or services as mail, railways and roads are examples of transportation networks. The optimal design fixes network architecture, including clustering, degree distribution, hierarchy, community structures and other structural metrics. These networks are specifically designed for efficient transportation, minimizing transit times and costs. All sorts of transportation networks face the same problem: traffic congestion among their channels. In this work we considered a transportation network model in which we optimize/minimize a cost function for the flux/current at each channel/link of the network. We performed simulations and an analytical study of this problem, focusing on the fraction of used channels and the flow distribution through these channels. Our results show that, after the initial transient, the fraction of used channels stays constant and, remarkably, this result does not depend on the lattice structure (2D, 3D, or long-range connections). For the case of high flow, all channels in the network are used. On the other hand, in the small flow limit, we observe a novel behavior that the fraction of used channels depends on the square root of the flow.

\end{abstract}


\doublespacing
\section{Introduction}

Network efficiency is a topic of great importance in network research, especially for distribution, delivery and transportation networks. Its main goal is the development of optimal designs, in order to build more effective connections, besides lower costs and transit time. The network's requirements determine its architecture, which is reflected in its clustering, degree distribution, hierarchic and community structures and other structural metrics. In our article, we consider transportation networks in which generalized flows are running on it. The optimization of these flows can be directly applied to any network-distributed goods, such as electricity, water, gas, telephone, data and traffic. 

All sorts of transportation networks face the same issue: traffic congestion in their channels. The traffic and its dynamics have been extensively studied in different areas, such as information routing ~\cite{Johnson, Royer} and water drainage in river basins~\cite{Rinaldo, Seybol, Edmonds}. It also has been studied by physicists~\cite{Helbing, sole, takayasu2, takayasu3} and, increasingly, by theoretical researchers~\cite{Banavar, martinho1, martinho2, xue, pastore, yang}.

Let us consider a transportation network with $N$ channels. The current $j$ flows through the network channels (links, bonds), between the intersections in the network (in other words, through the nodes), satisfying the flow conservation rule at each intersection, with $j_i \geq 0$. The transportation cost through the channels is usually related to the time required to transport goods to their destination. Considering that, one can write the total transportation cost $C$ as

\begin{equation}
 C = \sum_i e_i \left( a\cdot j_i + b\cdot j_i^2 \right),
\label{custo_total}
\end{equation}
where $e_i$ is a positive coefficient associated with each channel of the network and $A$ and $B$ are coefficients. Here we have neglected the higher order terms in the Eq.~\ref{custo_total} and, for convenience, we considered $a=1$ and $b=1/2$. 

When the input current is small, the optimal flow runs through a single chain of links with lower costs. When the input current increases, the optimal flow splits and the channels with higher costs are used. The resulting distribution of flows over links has the minimal value of $C$. One can determine the optimal current configuration among the channels by minimizing the cost function.

\begin{figure}
 \centering
 \resizebox{5cm}{!}{\includegraphics[width=\linewidth]{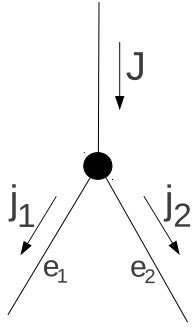}}
 \caption[Distribution of currents within two outgoing channels of a node.]{Distribution of currents within two outgoing channels of a node. The input current $J$ is divided in two, $j_1$ and $j_2$, associated with costs $e_1$ and $e_2$. When two channels meet at the same node their currents are simply summed into J.}
 \label{fig_fluxo}
\end{figure}

Considering the simplest case, a single node with an input current $J$ is connected with two outgoing channels, $j_1$ and $j_2$, see~Fig.~\ref{fig_fluxo}. Here, we consider a local optimization, with independent nodes, and the current flows in just one direction. For this simple case, one can write the cost function as

\begin{equation}
 C = e_1\left(j_1 + \frac{1}{2}j_1^2\right) + e_2\left(j_2 + \frac{1}{2}j_2^2\right).
 \label{cost}
\end{equation}

By using the current flow conservation rule, $j_1+j_2=J$, we can minimize the cost function, $\frac{\partial C}{\partial j_i} = 0$. The current flow is

\begin{equation}
 j_1 = \frac{e_2(J+1) - e_1}{e_1 + e_2}
 \label{flow1st}
\end{equation}
and
\begin{equation}
 j_2 = \frac{e_1(J+1) - e_2}{e_1 + e_2}.
 \label{flow2nd}
\end{equation}

These solutions permit negative current flows. Since we consider only positive currents, we must discard negative solutions. We can write the minimum input current $I_c$ as $I_c = e_2/e_1 - 1$. We can see that, for small input currents $J<I_c$, only one outgoing channel will be used (the one with minimal cost). On the other hand, for $J>I_c$, both outgoing channels will be used, minimizing the cost function.

\section{Mean-Field approach}

For a infinite system one should considered a mean-field version of this model. For that case, the nodes are also organized by layers and each node has two outgoing channels but the channels are randomly selected in pairs from the next layer, suppressing local correlations and changing the systems dimension to 1 + $\infty$. The structure is composed by an infinite number of consecutive layers of nodes, through which the current passes only once, without any loops on the network. In our model this dynamics is equivalent to the multilayer structure.

The flow distribution in the first layer depends on the initial distribution $P(j,t=0)$ and in the second layer depends on previously layer, and consequently on the initial distribution as well, and so on for the next layers. For the case when t $\rightarrow \infty$ the stationary current distribution is independent of initial conditions, with exception for the initial average current.

Simulations were supported by a mean-field theory~\cite{Rui-Orahcio,Orahcio} which gives

\begin{equation}
 1 - B = 2\sqrt{\langle j\rangle},
\end{equation}
and
\begin{equation}
 P(j)=4e^{\frac{-2j}{\sqrt{\langle j\rangle}}} 
\end{equation}
for small $\langle j \rangle$, as it is shown in Fig~\ref{fig_P(j)s}. Our simulations demonstrate that these laws work also in 2D and 3D and for large $\langle j \rangle$, beyond the limits of the applicability of any MF theory.

The minimal cost condition (eqs.~\ref{flow1st} and~\ref{flow2nd}) imposes a conditional probability $\mathcal{F}(j|J,I_c)$ for the output current $j$ given an input flow $J$ and the minimum value $I_c$,

\begin{eqnarray}
\mathcal{F}(j|J,I_c)  & = &  \left[ \frac{1}{2}\delta(j) +\frac{1}{2}\delta(j-J)\right]\Theta(I_c-J)+\nonumber\\
 & & + \left[\frac{1}{2}\delta\left(j-\frac{J-I_c}{2}\right)\right. +\nonumber\\
 & & \left. +\frac{1}{2}\delta\left(j-\frac{J+I_c}{2}\right)\right]\Theta(J-I_c)
\label{deltas}
\end{eqnarray}
where $\delta(j)$ and $\Theta(j)$ are the delta and step distributions respectively.

Since $\mathcal{P}_t(j)$ and $\mathcal{Q}_t(j)$ are the current distributions on the channels and the vertexes respectively at discrete time $t$, we can write a set of recursive integral equations (see appendix) for these distributions and the empty channels probability $B_t$, straightly defined by the $\mathcal{P}_t(j)$ distribution
\begin{equation}
B_t = 1 - \int_0^\infty \mathcal{P}_t(j)dj.
\end{equation}

The cost distribution -- or the critical current distribution $v(I_c)$ -- is an important factor on the final form of the above distributions. We propose a simple shape for that distribution to perform the mean field approach,
\begin{equation}
v(I_c) = \frac{1}{\beta^2}\Theta(\beta-I_c)
\end{equation}
is the uniform case, where the critical flow are uniformly distributed on the net, therefore, $\beta$ is the maximum critical current found.

The solution of the recurrence equations set can be an exhaustive task even for computers, each point in the next distribution depends on the integration of thousand points in the previous distribution. But that set of equations allows some limit approaches for steady terms like the empty channel probability $B$ for small currents, the coefficient $\mathcal{P}(0)$ for the current channel distribution, still on the small current limit and the large current limit distribution.

On the small current limit, the uniform case gives a steady equation for the $B$ probability in the form
\begin{equation}
B(\langle j\rangle,\langle j^2\rangle_\mathcal{Q}) = 1-2\sqrt{\langle j\rangle - \frac{\langle j^2\rangle_\mathcal{Q}}{4}},
\end{equation}
where we choose $\beta=1$ and the $\langle\rangle$ denotes a moment performed by the $\mathcal{P}(j)$ distribution and the $\langle\rangle_\mathcal{Q}$ for the $\mathcal{Q}(j)$ one. The numerical integration of the \ref{generalrec} equations with $t_{max}=250$ time steps, allow us the strong assumption for the second moment $\langle j^2\rangle_\mathcal{Q}$, it decays to zero faster than the average current $\langle j\rangle$. So the probability of active channels is $1-B\sim\langle j\rangle ^{1/2}$ on this limit.

By an analog way, the steady coefficient $\mathcal{P}(0)$ can be evaluated with the \emph{posteriori} current distribution on the \ref{generalrec} equations,
\begin{equation}
\mathcal{P}(0) = 2(1+B)-4\frac{\langle j\rangle}{1-B}.
\end{equation}
Since $B\rightarrow 1$ for the small current limit and $1-B$ goes to zero slower than the average current, according the previous assumption, thus the steady coefficient $P(0)=~4$.

For the large currents limit $\langle j\rangle\geqslant\beta$, we can employ Laplace transform technique~\cite{ross} on recurrence equations~\ref{generalrec} and~\ref{generalrec2}. Since $\Pi(z)$ and $K(z)$ are the transforms for $\mathcal{P}(j)$ and $\mathcal{Q}(j)$ distributions respectively, the Laplace recurrence equations become 
\begin{equation}
\Pi_{t+1}(z) = \left[\frac{4}{\beta z}\sinh\left(\frac{\beta z}{4}\right) \Pi_t\left(\frac{z}{2} \right)\right]^2.
\label{largeite}
\end{equation}

If an initial distribution for channels is proposed, there will be a final form for the limit of infinite iterations on the above equation~\ref{largeite}. After $n$ iterations it can be able to invert analytically that equation an perform numerically the inset plot on figure~\ref{fig_P(j)}.

The shape of the above distribution is close to a normal one, but by evaluation of each moment generating function derivatives, equation~\ref{largeite}, on the $z\rightarrow 0$ limit, we conclude that the steady distribution has just three moments equal to the normal distribution ones: the mean, the variance and the third moment. Table~\ref{tab:moments} compares first five non null central moments between the steady distribution and the normal one.

\begin{table}[ht]
\centering
\caption{\label{tab:moments} Top six non-null moments of normal (A) and the steady distribution for large currents from this paper (B). Variance $\sigma^2=\frac{\beta^2}{12}$.}
\begin{tabular}{@{}lccccc@{}}
\hline
\textbf{Order}      & \textbf{2nd}        & \textbf{4th}                         & \textbf{6th}                            & \textbf{8th}                                   & \textbf{10th}                                       \\ \hline
\textbf{A}     & $\sigma^2$ & $3\sigma^4$                 & $15\sigma^6$                   & $105\sigma^8$                         & $945\sigma^{10}$                              \\
\textbf{B} & $\sigma^2$ & $\frac{99}{35}\sigma^4$ & $\frac{2745}{217}\sigma^6$ & $\frac{10409283}{137795}\sigma^8$ & $\frac{12298729365}{22129877}\sigma^{10}$ \\ \hline
\end{tabular}
\end{table}
\section{Simulations and Results}

\begin{figure}
\vspace{0.25cm}
 \includegraphics[width=.4\textwidth]{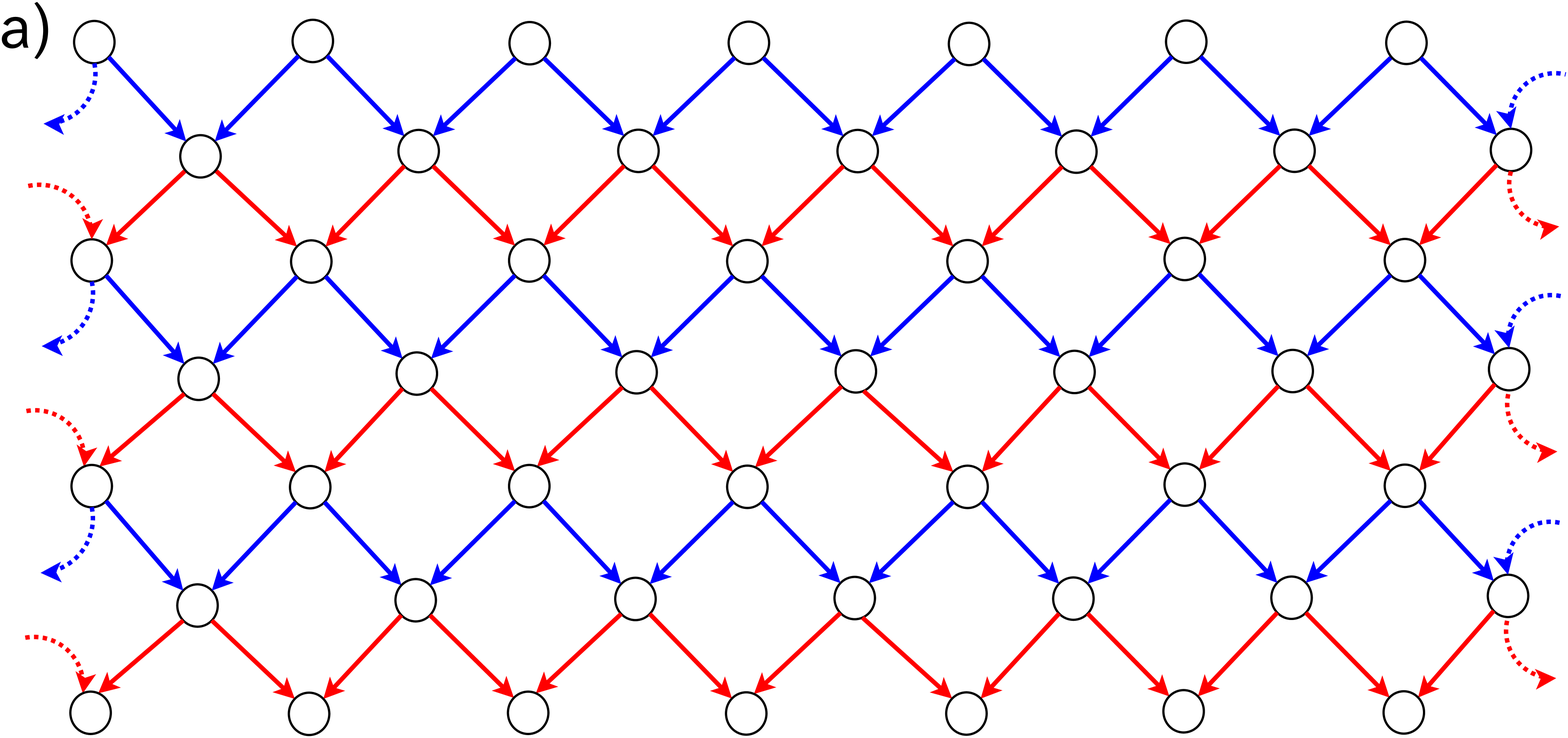}\hfill
 \includegraphics[width=.3\textwidth]{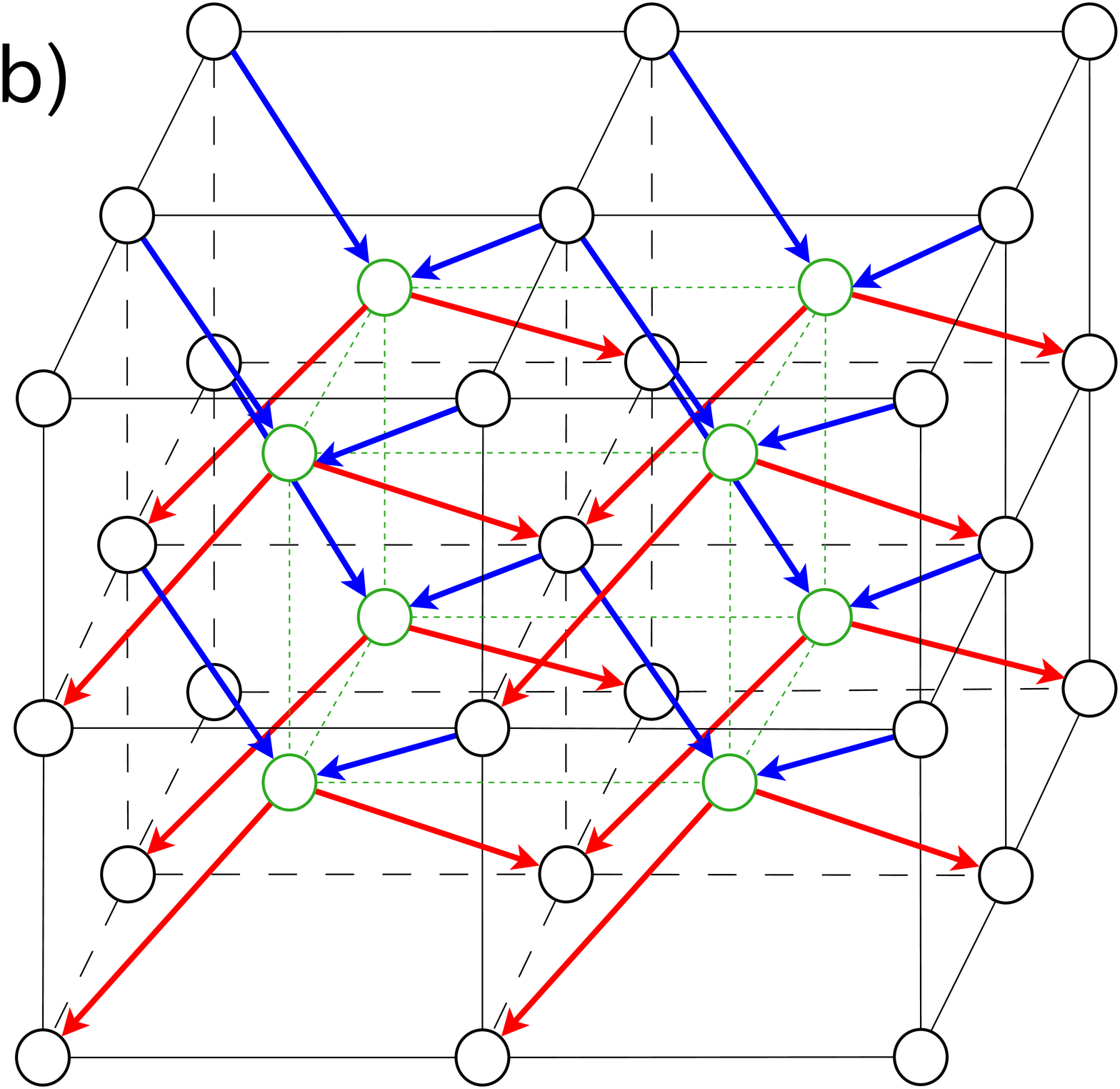}\hfill
 \caption[Simulation for optimized ``lattices'' in the flow model.]{The ``lattices'' for the simulations in the flow optimization model. Each node is connected with four directed channels (links), two from the top and two from the bottom layer, both for the simulations with two dimension (a) and three dimensions (b). We used periodic boundary conditions in our simulations.}
 \label{fig_celula}
\end{figure}

Depending on the input flow, a smaller or greater fraction of the network is used, so the quantity of interest on this problem is the number of used channels, i.e. with current running through them. One can perform computer simulations and measure the number of empty channels $B$. We consider a directed network, with four channels (two incoming and two outgoing) for each node. In our simulations we considered three different situations: two-dimensional lattices, three-dimensional lattices and mean-field (infinite long-range connections) cases. For all cases the current flows from top to bottom (Fig.~\ref{fig_celula}). For the mean-field case simulations, we have considered randomly connected sites in the neighboring that are also connected uniformly at random, so each site is connected to two randomly chosen sites from the previous layer.

In our simulations we locally optimize the current flow. At each node, the currents from two incoming channels are summed. This new current $J$ is then divided into the two outgoing currents $j_1$ and $j_2$, as shown in Fig.~\ref{fig_fluxo}. If $J<I_c$, the current will flow through just one outgoing channel. Otherwise, if $J>I_c$, the current will flow through both outgoing channels. At first, the current flow for the entire layer is optimized, so time corresponds to the $t^{th}$ layer. Note that the total current $N \times \langle j \rangle$ is conserved, i.e., it is the same for every layer.

\begin{figure}
 \centering
 \includegraphics[width=0.9\linewidth]{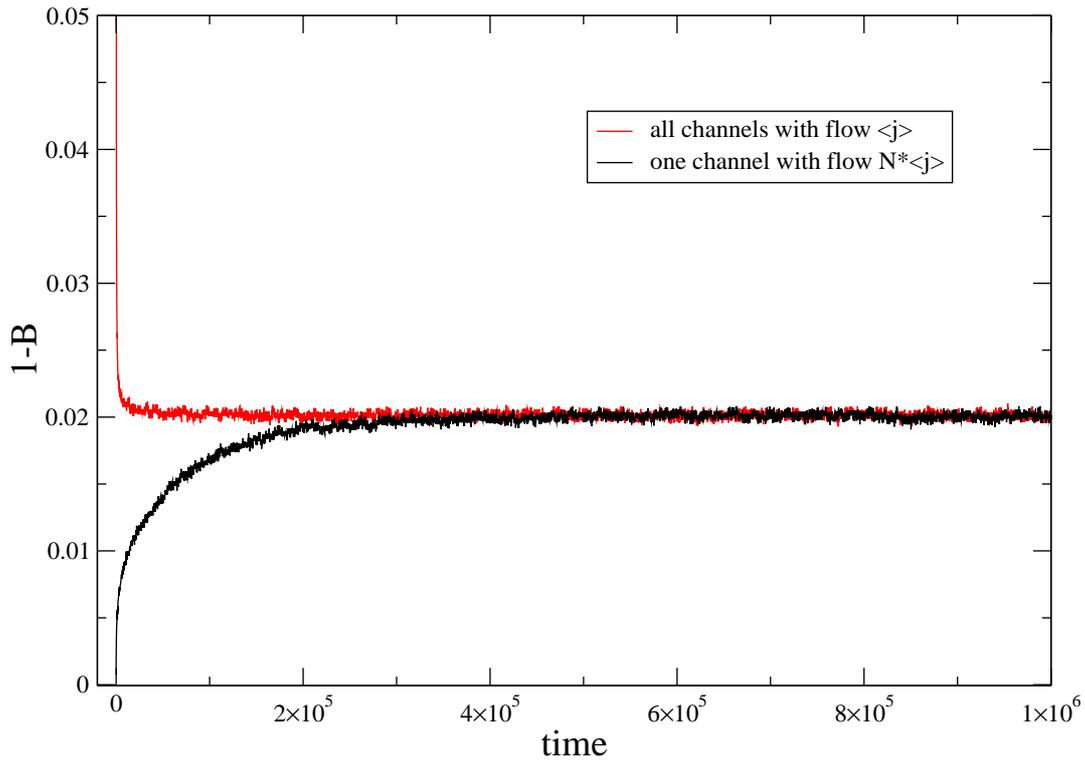}
 \caption[Two different initial configurations for the current flows in 2D.]{Two different initial configurations for the current flow in 2D. Using the same total current flow, in the first case (red line) the total current is equally divided for all channels. For the second case (black line), the total current is initially introduced in only one channel. For both cases the results are for 1000 channels, $\langle j \rangle = 10^{-4}$ and averaged over 100 samples. One can see that, despite having different relaxation times, both situations reach the same steady state with the same fraction of used channels.}
 \label{fig_1-B_diff_ini_cond}
\end{figure}

We start our simulations by injecting a total current $N \times \langle j \rangle$ in the first layer, when the costs of each channel are uniformly distributed in the interval $0\leq e_i \le 1$. The fraction of used channels $1-B$ as a function of time, i.e., the number of the current layer, is shown in Fig.~\ref{fig_1-B_diff_ini_cond}, for the two dimensional simulations, using $N=1000$, $\langle j \rangle = 10^{-4}$ and averaged over 100 samples.

In our simulations we have used two different initial configurations. In the first one, we set the total input current equally divided between all channels. In the second one, we set the total current only in one channel. As one can see, the initial configuration is not important for the stationary regime, since after relaxation both configurations display the same result.

\begin{figure}
 \centering
 \includegraphics[width=0.9\linewidth]{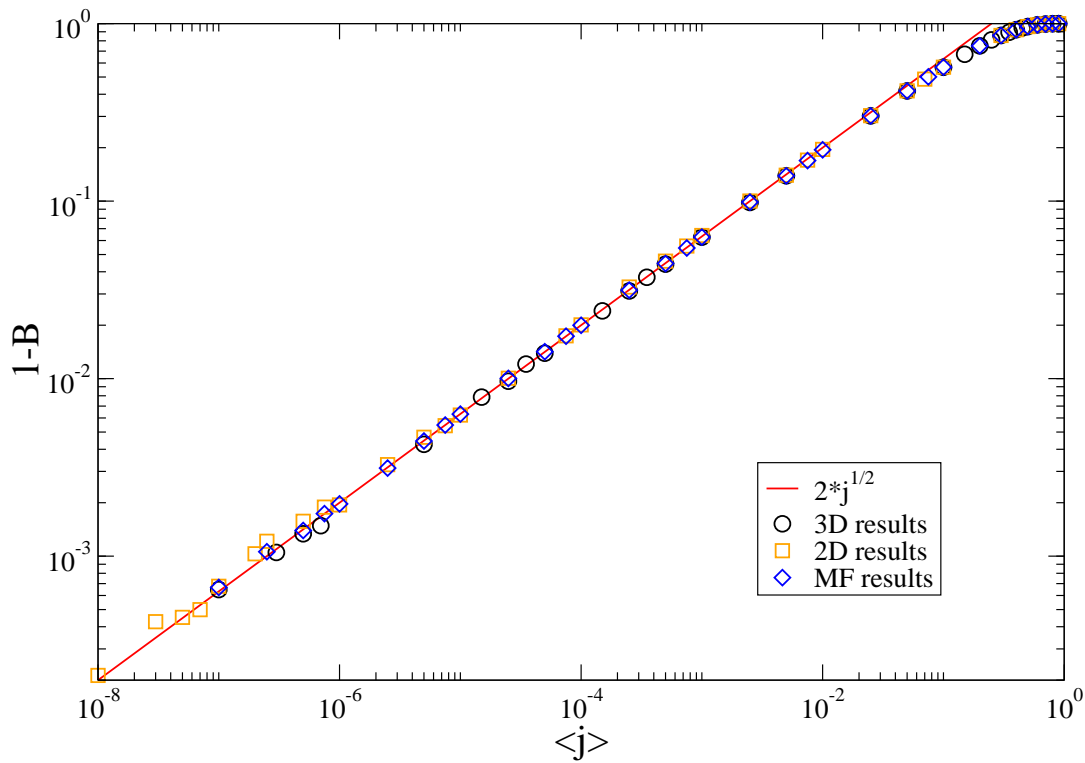}
 \caption[Fraction of used channels as function of $\langle j\rangle$ for mean-field, 2- and 3-D results.]{Fraction of used channels as function of $\langle j\rangle$ for mean-field, two- and three-dimensional results. One can see that 2D, 3D and mean-field networks provide the same stationary results, following the $2\langle j\rangle^{\frac{1}{2}}$ law in the limit of small current (red straight line).}
 \label{fig_1-B}
\end{figure}

After the initial transient, the fraction of used channels $1-B$ on the network stays constant. One can plot $(1-B)$ at steady state as function of $\langle j \rangle$. Remarkably, this result does not depend on the lattice (2D, 3D, or long-range connections), as one can see in Fig.~\ref{fig_1-B}. All the configurations show the same result, for a wide range of $\langle j \rangle$ in the small currents limit.

For the case of the high current limit, where $\langle j \rangle \rightarrow 1$, all channels on the network are used. In the small current limit, $\langle j \rangle \ll 1$, we observe that the fraction of used channels depends on $(1-B) \sim 2j^{\frac{1}{2}}$.

\begin{figure}
 \vspace{0.5cm}
 \centering
 \includegraphics[width=0.9\linewidth]{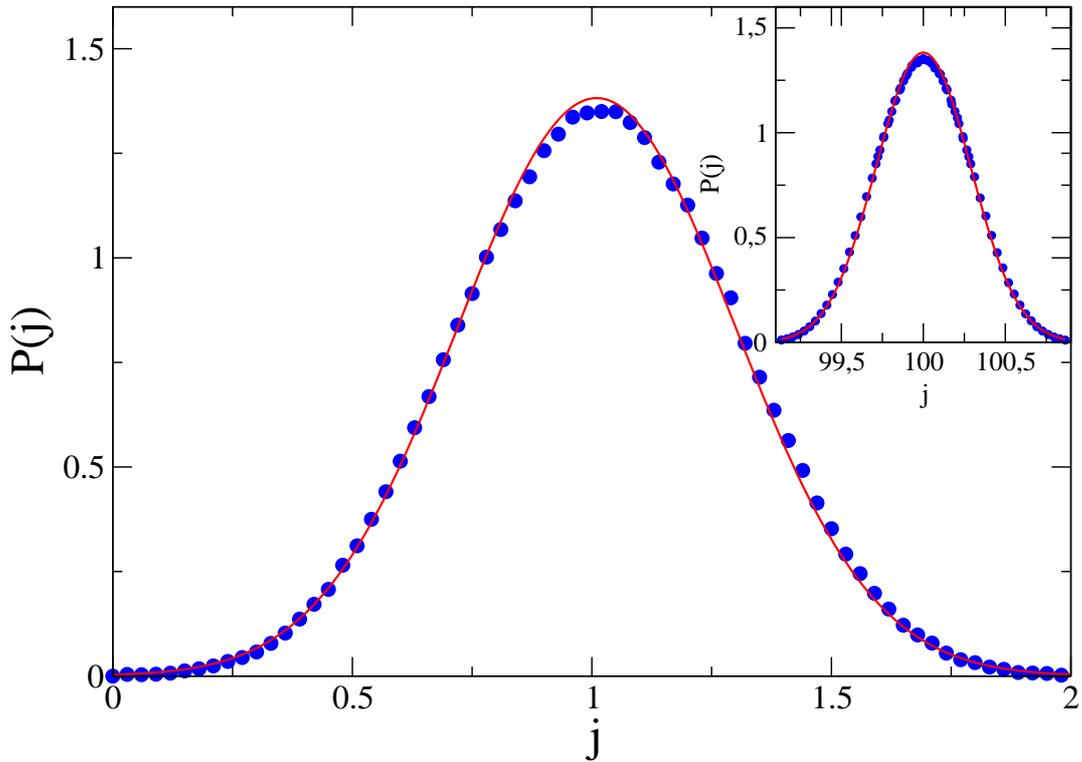}
 \caption[The current distribution for a channel in the limit of large input currents.]{The current distribution for currents on channels in the limit of large input currents, in this case, $\langle j\rangle = 1$. The points are the result of our simulations and the red straight line is a Gaussian fit with $\mu = 1.01$ (mean) and $\sigma^2 = 0.0872 \simeq \frac{1}{12}$ (variance). Inset is the numerical calculation from analytical solution for $\langle j\rangle=100$, again the red straight line corresponds to a Gaussian curve with $\mu=100$ and $\sigma^2=\frac{1}{12}$.}
 \label{fig_P(j)}
\end{figure}

\begin{figure*}[!htb]
\vspace{0.25cm}
\includegraphics[width=.33\textwidth]{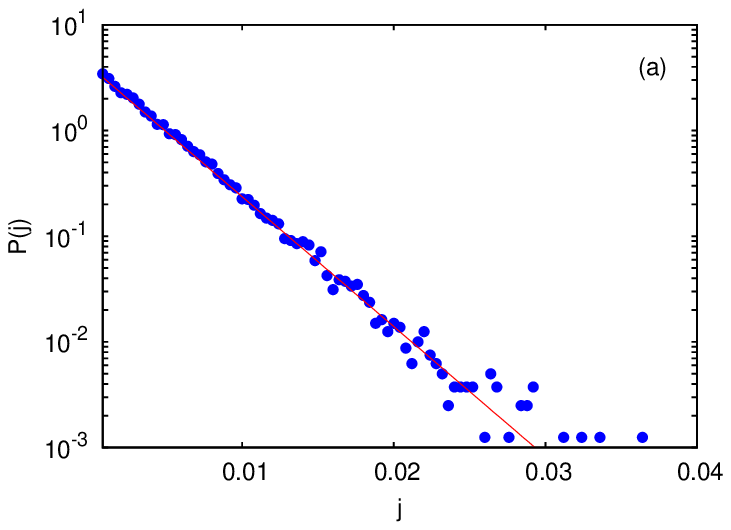}\hfill
\includegraphics[width=.33\textwidth]{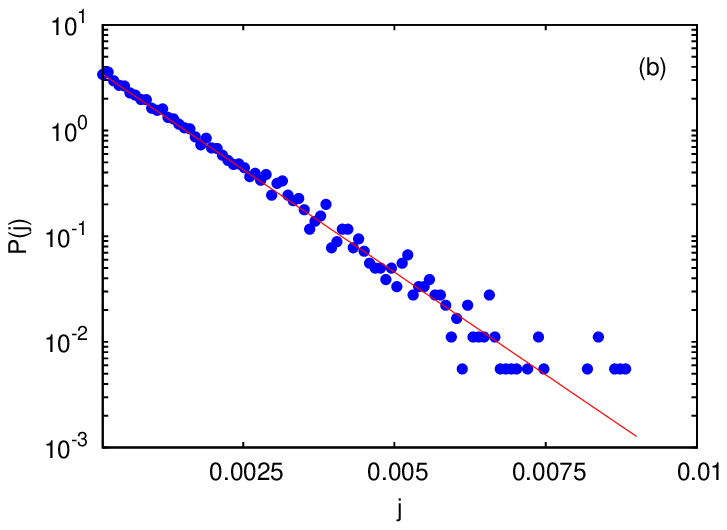}\hfill
\includegraphics[width=.33\textwidth]{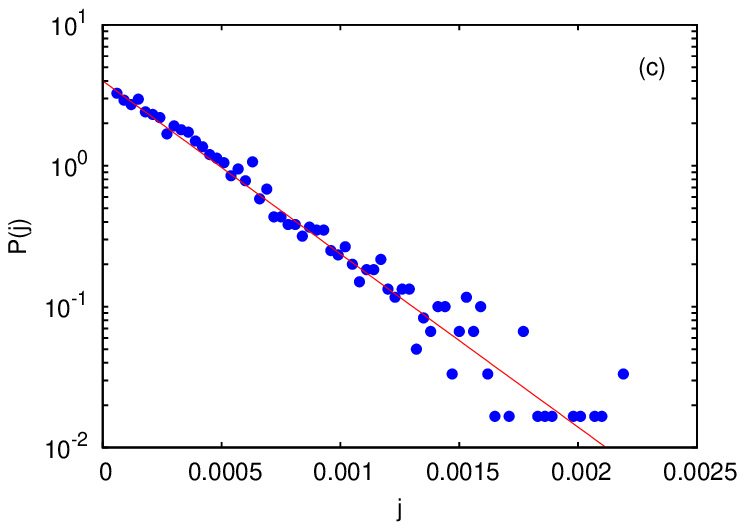}\hfill
\caption[The current distribution in the limit of small current flows.]{The current distribution in the limit of small current flows for different values of the $\langle j\rangle$. The straight line is the asymptotic value $P(j)=4e^{\frac{-2j}{\sqrt{\langle j\rangle}}}$ and the points are the results from our simulations for $\langle j\rangle=5 \times 10^{-5}$ (a), $\langle j\rangle=5 \times 10^{-6}$ (b), and $\langle j\rangle=5 \times 10^{-7}$ (c).}
\label{fig_P(j)s}
\end{figure*}

From our simulations, we obtained the distribution of the currents, $P(j)$. For large input currents, $\langle j \rangle \sim 1$, $P(j)$ follows a Gaussian distribution, as in Fig.~\ref{fig_P(j)}. The best fit with the Gaussian distribution is obtained for the values of $\mu = 1.01$ and $\sigma^2 = 0.0872 \simeq \frac{1}{12}$. On the other hand, when we consider the limit of the small current flows, $\langle j \rangle \ll 1$, we found that the current distribution has an exponential dependence with $\langle j \rangle^{-\frac{1}{2}}$, as shown in Fig.~\ref{fig_P(j)s}.

\section{Conclusion and Remarks}

We studied a general transportation network in which current flows through the network's channels, with randomness introduced by a random cost function in the channels. In our flow optimization model, in which the current flows through a random network that is actually a lattice, the randomness is due to random coefficients of a cost function defined at the lattice bonds. We find that the stationary flow distribution strongly depends on the amount of flow initially injected in the system and it is independent of the topological structure. The average current per node is the only independent parameter of the model. We obtained the exponential current distribution for small currents limits. For the large ones, the distribution is different from a Gaussian distribution but equally centered on the mean and with the same variance, i.e., fluctuations on large flows limit depends on the distribution of critical flows (or costs) not on the average flow. We found that if $\langle j \rangle$ is small, all the currents flow through a tiny fraction of the channels by a power-law dependence (with mean-field exponent $\frac{1}{2}$) of the fraction of used channels with the mean input current $\langle j \rangle$, and that MF describes even 2D and 3D cases.

\section*{Acknowledgments}
F. L. Forgerini would like to thank the FCT for the financial support by project No. SFRH/BD/68813/2010. And O. F. de Sousa would like to thanks the financial support from FAPERJ by \emph{Estágio de doutorandos no exterior 2011/1} program.
\newline

\appendix*
\section{Mean field recurrence equations}

The integral recurrence equations of the current distributions on vertexes $\mathcal{Q}_t(j)$, on channels $\mathcal{P}_t(j)$ and the fraction of empty vertexes $B_t$ are given by 

\begin{eqnarray}
\label{generalrec}
\mathcal{Q}_t(j) & = 2B_t \mathcal{P}_t(j) + \int_0^j \mathcal{P}_t(j)\mathcal{P}_t(j-J) dJ \nonumber \\
\label{generalrec2}
\mathcal{P}_{t+1}(j) & = \frac{\mathcal{V}(j)}{2}\mathcal{Q}_t(j) + 2\int_0^\infty v(J)\mathcal{Q}_t(2j+J) dJ + 2 \int_0^j v(J)\mathcal{Q}_t(2j-J) dJ \nonumber \\
B_{t+1} & = B_t^2+ \frac{1}{2}\int_0^\infty \mathcal{V}(j)\mathcal{Q}_t(j) dj \nonumber
\end{eqnarray}
\noindent where $\mathcal{V}(j)$ is two times the complementary cumulative distribution of critical currents distribution $v(j)$ defined by the channel costs, the $2$ factor is due the costs values degenerescence by two channels.

The set of equations above follow the current split rule on a bifurcation: (i) in time $t$, $\mathcal{Q}_t$ has two terms, the first one is the empty channel plus filled one by a current $j$ and the second term is the coalescence of two filled channels; (ii) next step ($t+1$) one random channel receives a current $j$ if this one is bellow the critical current defined by the $v(j)$ distribution, first term or one channel probability, otherwise the $j$ value will depend on the current split probability, second and third terms; (iii) finally the empty channel probability on the next step depends on two empty channels, first term, and one filled channel joins to a empty one, second term.

\end{document}